\documentclass[prd,showpacs,superscriptaddress,nofootinbib,floatfix,11pt]{revtex4-2}
\usepackage[utf8]{inputenc}
\usepackage{amsmath,amssymb}
\usepackage{slashed}
\usepackage{subfigure}
\usepackage[colorlinks=true,
breaklinks=true,
urlcolor=blue,
citecolor=blue]{hyperref}
\usepackage[usenames,dvipsnames]{color}
\usepackage{appendix}
\usepackage{braket,bm}
\usepackage{multirow}
\usepackage{enumitem}
\usepackage{color}
\usepackage{slashed}
\usepackage{orcidlink}
\usepackage{graphicx}
\usepackage{tikz}
\usepackage{booktabs}
\usepackage{array}
\usepackage{overpic}
\usepackage{float}
\usepackage{threeparttable}
\allowdisplaybreaks[4]

\newcommand{\zjwl}{\affiliation{College of Information and Intelligence Engineering, Zhejiang Wanli University, Zhejiang 315101, China}}

\newcommand{\nbu}{\affiliation{Physics Department, Ningbo University, Zhejiang 315211, China}}

\newcommand{\kmu}{\affiliation{School of Physics Science and Technology, Kunming University, Kunming 650214, China}}

\newcommand{\usc}{\affiliation{School of Nuclear Science and Technology, University of South China, Hengyang, 421001, Hunan, China}}

\begin{document}
\title{Production of $D_s\bar{D}_s$ and $D\bar{D}$ bound states in the $B$ decays within the Bethe–Salpeter framework} 
\author{Zhen-Yang Wang\orcidlink{0000-0002-4074-7892}}\email{wangzhenyang@nbu.edu.cn}
\nbu
\author{Jing-Juan Qi\orcidlink{0000-0002-9260-9408}}\email{qijj@mail.bnu.edu.cn}
\zjwl\nbu
\author{Zhen-Hua Zhang\orcidlink{0000-0001-5031-9499}}\email{zhangzh@usc.edu.cn}
\usc
\author{Xin-Heng Guo\orcidlink{0000-0002-9309-9112}}\email{corresponding author:xhguo@bnu.edu.cn}
\kmu

\begin{abstract}
Within the Bethe--Salpeter framework, we investigate the production of possible $D_s\bar{D}_s$ $(X_{s\bar{s}})$ and $D\bar{D}$ $(X_{q\bar{q}})$ bound states in $B$ decays. The bound state properties of the two heavy meson systems are studied in the one-boson-exchange model, and the resulting normalized Bethe--Salpeter wave functions are used to calculate the branching fractions of $B^+\to X_{s\bar s}K^+$ and $B^+\to X_{q\bar q}K^+$. We find that bound state solutions for the $D\bar{D}$ system exist for all the three coupling sets considered, whereas the $D_s\bar{D}_s$ system supports a bound-state solution only in a restricted parameter region. The predicted branching fractions are in the ranges of $1.09\times10^{-5}$--$20.06\times10^{-4}$ for the $D_s\bar{D}_s$ bound state and $1.56\times10^{-6}$--$4.14\times10^{-4}$ for the $D\bar{D}$ bound state. In particular, if $X(3915)$ is interpreted as a predominantly $D_s\bar{D}_s$ bound state, its production branching fraction in $B$ decay is found to be $5.79\times10^{-4}$.
\end{abstract}
	
	\date{\today}	
	\maketitle
\newpage

\section{Introduction}\label{sec1}
In 1964, Gell-Mann \cite{Gell-Mann:1964ewy} and Zweig \cite{Zweig:1964ruk} independently proposed the quark model, in which conventional hadrons are classified into mesons composed of a quark and an antiquark ($q\bar{q}$) and baryons composed of three quarks ($qqq$). Beyond these conventional configurations, quantum chromodynamics (QCD) also allows for more complicated hadron structures, such as tetraquark states, pentaquark states, hexaquark states, glueballs, and hybrid states, which are collectively referred to as exotic hadrons. In 2003, the Belle Collaboration reported the $X(3872)$ in the exclusive decay process $B^+\rightarrow K^+\pi^+\pi^-J/\psi$, which has since become one of the most prominent candidates for exotic hadrons \cite{Belle:2003nnu}. Since then, numerous exotic candidates have been observed experimentally, including $Y(4260)$, $Z_c(3900)$, $Z_{cs}(3985)$, $P_c$, $T_{cc}$, $T_{cs}$, and $X(6900)$ \cite{ParticleDataGroup:2024cfk}. A notable feature of many of these states is that their masses are located close to the thresholds of two conventional hadrons, which suggests a possible hadronic molecule interpretation, namely loose bound states generated by deuteronlike meson-exchange forces. The study of exotic hadrons has attracted extensive attention from both experimentalists and theorists, providing valuable insight into hadron spectroscopy and nonperturbative QCD.

The LHCb Collaboration reported the $X(3960)$ in the decay $B^+\rightarrow D_s^+D_s^-K^+$ \cite{LHCb:2022aki} and the $\chi_{c0}(3930)$ in $B^+\rightarrow D^+D^-K^+$ \cite{LHCb:2020pxc}. Since both structures lie close to the $D_s\bar{D}_s$ threshold, their nature has attracted considerable attention. The Particle Data Group (PDG) currently assigns both of them to $X(3915)$ with $I(J^{PC})=0(0^{++})$, which was first observed by the Belle Collaboration in the $J/\psi\omega$ invariant mass spectrum \cite{Belle:2004lle}. On the theoretical side, the $D_s^+D_s^-$ invariant mass distribution measured in $B^+\rightarrow D_s^+D_s^-K^+$ can be described within nonrelativistic effective field theory by introducing a $D_s^+D_s^-$ bound or virtual state below threshold \cite{Ji:2022uie}. In addition, a coupled-channel study of the $D\bar{D}$ and $D_s\bar{D}_s$ interactions in $B^+\rightarrow D^+D^-K^+$ found two bound states below the $D_s\bar{D}_s$ and $D\bar{D}$ thresholds, respectively, and reproduced the near-threshold enhancement in the $D_s^+D_s^-$ mass distribution, which is in agreement with the LHCb observation \cite{Bayar:2022dqa}. These results suggest that the near-threshold enhancement can be understood without introducing an additional $X(3960)$ state and is compatible with the current PDG assignment.

The possible existence of $D_s\bar{D}_s$ and $D\bar{D}$ bound states has been investigated using a variety of theoretical approaches. Lattice QCD studies of coupled-channel $D\bar{D}$ and $D_s\bar{D}_s$ scattering  predicted a $D\bar{D}$ bound state and a narrow $0^{++}$ resonance slightly below the $D_s\bar{D}_s$ threshold \cite{Prelovsek:2020eiw}. Related studies have also been carried out within the contact-range theory \cite{Peng:2023lfw}, the vector-meson-dominance model \cite{Dong:2021juy}, the one-boson-exchange model \cite{Ding:2021igr,Liu:2017mrh,Ke:2022vsi}, the Lippmann-Schwinger equations \cite{Meng:2020cbk}, the coupled-channel Bethe-Salpeter (BS) equations \cite{Abreu:2025jqy}, and the unitarized coupled-channel framework \cite{Gamermann:2006nm}. Most theoretical studies support the existence of a $D\bar{D}$ bound state, whereas the existence of a $D_s\bar{D}_s$ bound state remains unsettled.

In addition, the production rates of $D_s\bar{D}_s$ and $D\bar{D}$ molecules in $B$ and $B_s$ decays have been discussed in Refs.~\cite{Xie:2022lyw,Wu:2023fyh}. The $D\bar{D}$ invariant mass distributions in the processes $\gamma\gamma\rightarrow D\bar{D}$ \cite{Wang:2020elp,Deineka:2021aeu}, $B\rightarrow D\bar{D}K$ \cite{Dai:2015bcc}, $\psi(3770) \rightarrow \gamma D\bar{D}$ \cite{Dai:2020yfu}, and $\Lambda_b\rightarrow \Lambda D\bar{D}$ \cite{Wei:2021usz}, as well as the $\eta\eta_c$ invariant mass distribution in the decay $B^-\rightarrow K^-\eta\eta_c$ \cite{Li:2023nsw}, have been analyzed in connection with these bound state interpretations. The coupled-channel interaction between $D\bar{D}$ and $D_s\bar{D}_s$ in the decay $B^-\rightarrow K^-J/\psi\omega$ has also been investigated \cite{Abreu:2023rye}. Moreover, the decay properties of $X(3915)$ as a $D_s\bar{D}_s$ bound state have been studied in Refs. \cite{Li:2015iga,Qi:2023gwb}. Nevertheless, whether $\chi_{c0}(3930)$ can be identified as a $D_s\bar{D}_s$ bound state remains unsettled, and no experimental candidate for a $D\bar{D}$ bound state has yet been firmly established. Further investigations of these states are therefore highly needed.

The decay properties of the $S$-wave $D_s\bar{D}_s$ bound state were investigated in our previous work. In the present work, we extend our study to the production of $D_s\bar{D}_s$ and $D\bar{D}$ bound states in $B$ decays. Following the notation of Ref. \cite{Xie:2022lyw}, these two bound states are denoted by $X_{s\bar{s}}$ and $X_{q\bar{q}}$, respectively. By solving the BS equations with the vector-meson exchange kernel under the ladder and instantaneous approximations, we obtain the BS wave functions of the $D_s\bar{D}_s$ and $D\bar{D}$ bound states. These Lorentz-covariant BS wave functions include the essential internal dynamics of the bound states and are subsequently employed to calculate their production rates in $B$ decays. We further analyze how the bound state properties affect the predicted production rates, which may provide useful guidance for future experimental searches.

The remainder of this paper is organized as follows. In Sec. \ref{TF}, we present the amplitudes for $B^+\rightarrow X_{s\bar{s}}K^+ $ and $B^+\rightarrow X_{q\bar{q}}K^+$, together with the BS equations for the $S$-wave $D_s\bar{D}_s$ and $D\bar{D}$ bound states under the ladder and instantaneous approximations. The numerical solutions of the BS equations and the dependence of the production rates on the binding energy are presented in Sec. \ref{NR}. Our conclusions are given in the final section.

\section{Theoretical formalism}\label{TF}
\begin{figure}[ht]
\centering
    \includegraphics[width=0.9\textwidth]{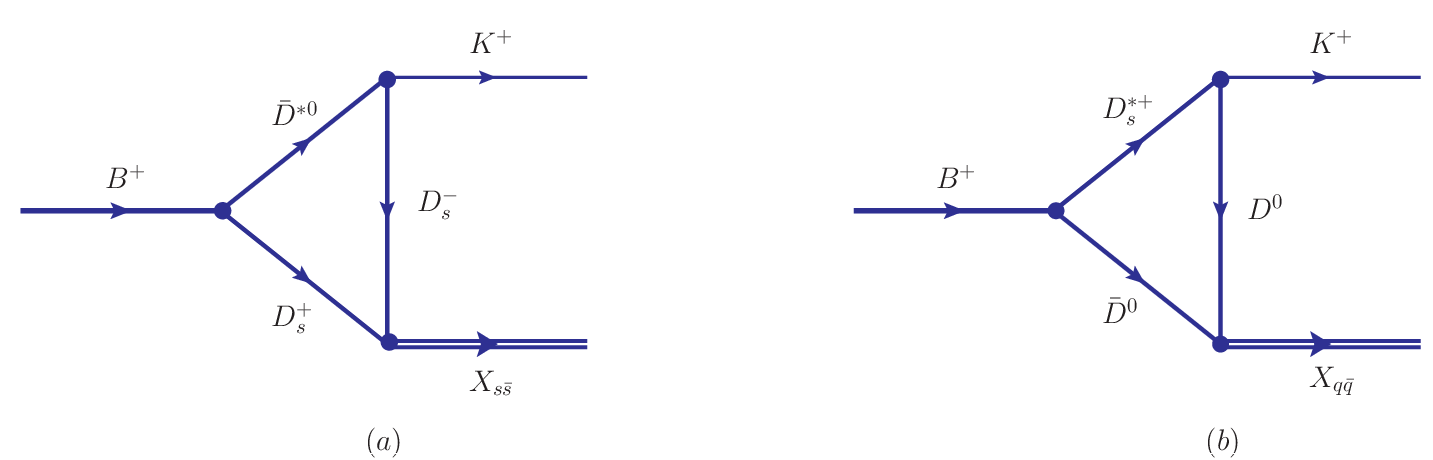}
    \caption{Diagrams for the hadron-level processes (a) $B^+\rightarrow X_{s\bar{s}}K^+$ and (b) $B^+\rightarrow X_{q\bar{q}}K^+$.}
  \label{fig1}
\end{figure}

In this work, we focus on the triangle-diagram contributions shown in Fig. \ref{fig1}, which are assumed to provide the dominant production mechanism for $X_{s\bar{s}}$ and $X_{q\bar{q}}$ in $B^+$ decays. In this mechanism, the $B$ meson first undergoes a weak decay into a pair of charmed mesons. The intermediate $\bar{D}^{\ast0}$ or $D_s^{\ast+}$ meson then decays strongly, and the remaining meson pair subsequently rescatters to form the bound state $X$, such a rescattering mechanism has been successfully employed in the studies of exotic hadron production in $B$ decays and other heavy-flavor processes \cite{Liu:2022dmm,Wu:2023fyh,Oset:2024fbk,Yu:2023nmb,Nakamura:2022gtu,Wu:2021caw}.

\subsection{Weak decays of the $B$ meson}

\begin{figure}[ht]
\centering
    \includegraphics[width=1\textwidth]{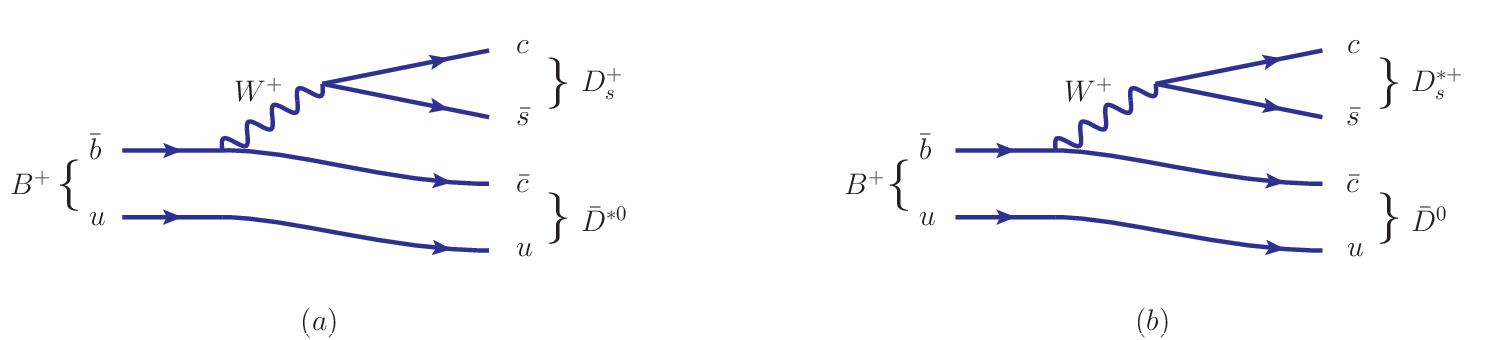}
    \caption{Quark-level diagrams for (a) $B^+\rightarrow D_s^+\bar{D}^{\ast0}$and (b) $B^+\rightarrow D_s^{\ast+}\bar{D}^{0}$.}
  \label{fig2}
\end{figure}

For the weak interaction part, the $B^+$ meson decays into a pair of charmed mesons through the processes $B^+\rightarrow D_s^{\ast+}\bar{D}^0$ and $B^+\rightarrow D_s^{+}\bar{D}^{\ast0}$. At the quark level, the $\bar{b}$ quark decays into a $\bar{c}$ quark by emitting a virtual $W^+$ boson, which subsequently decays into a $c\bar{s}$ quark pair. Both processes are Cabibbo favored, as depicted in Fig. \ref{fig2}.

Within the operator product expansion, the effective weak Hamiltonian relevant to the decays $B^+\rightarrow D_s^{\ast+}\bar{D}^0$ and $B^+\rightarrow D_s^+\bar{D}^{\ast0}$ can be written as \cite{Beneke:2003zv}
\begin{equation}
    \mathcal{H}{\text{eff}}=\frac{G_F}{\sqrt{2}}V_{cb}V_{cs}^\ast\left[c_1(\mu)Q_1+c_2(\mu)Q_2\right]+\text{H.c.},
\end{equation}
where $G_F$ is the Fermi constant, $V_{cb}$ and $V_{cs}$ are the Cabibbo–Kobayashi–Maskawa (CKM) matrix elements, and $c_{1,2}(\mu)$ are the Wilson coefficients at the renormalization scale $\mu$. Contributions from penguin operators are neglected because they are CKM suppressed in bottom-meson decays. The four-quark operators $Q_{1,2}$ are given by
\begin{equation}
\begin{split}
Q_1&=\left[\bar{c}_i\gamma_\mu(1-\gamma_5)s_i\right]\left[\bar{b}_j\gamma^\mu(1-\gamma_5)c_j\right],\\
Q_2&=\left[\bar{c}_i\gamma_\mu(1-\gamma_5)s_j\right]\left[\bar{b}_j\gamma^\mu(1-\gamma_5)c_i\right],
\end{split}
\end{equation}
where $s$, $c$, and $b$ denote the corresponding quark fields, and the subscripts $i$ and $j$ denote color indices. 

In the factorization approximation, the hadronic amplitudes for $B^+\rightarrow D_s^{\ast+}\bar{D}^0$ and $B^+\rightarrow D_s^+\bar{D}^{\ast0}$ are expressed as 
\begin{equation}
    \mathcal{A}(B^+\rightarrow D_s^+\bar{D}^{\ast0})=\frac{G_F}{\sqrt{2}}V_{cb}V_{cs}^\ast a_1\langle D_s^+|(\bar{c}s)|0\rangle\langle \bar{D}^{\ast0}|(\bar{b}c)|B^+\rangle,
\end{equation}
and
\begin{equation}
    \mathcal{A}(B^+\rightarrow D_s^{\ast+}\bar{D}^{0})=\frac{G_F}{\sqrt{2}}V_{cb}V_{cs}^\ast a_1'\langle D_s^{\ast+}|(\bar{c}s)|0\rangle\langle \bar{D}^{0}|(\bar{b}c)|B^+\rangle,
\end{equation}
where $(\bar{q}_1q_2)$ denotes the $V-A$ current $\bar{q}_1\gamma^\mu(1-\gamma_5)q_2$, and $a_1^{(')}=c_1+c_2/N_c$ is the effective Wilson coefficient in the naive factorization approach, with $N_c$ being the effective number of colors. 

The current matrix elements between the vacuum and the pseudoscalar or vector meson are parameterized as
\begin{equation}\label{vacuum vertex}
\begin{split}
       \langle D_s^+|(s\bar{c})|0\rangle&=f_{D_s^+}p^\mu_{D_s^+},\\ 
       \langle D_s^{\ast+}|(s\bar{c})|0\rangle&=m_{D_s^{\ast+}}f_{D_s^{\ast+}}\epsilon_\mu^\ast,
\end{split}
\end{equation}
where $f_{D_s^+}=250$ MeV and $f_{D_s^{\ast+}}=272$ MeV are the decay constants of $D_s^+$ and $D_s^{\ast+}$, respectively, and $\epsilon_\mu^\ast$ represents the polarization vector of $D_s^{\ast+}$.

The $B^+\to \bar{D}^{(\ast)0}$ transition matrix elements are parameterized in terms of form factors as \cite{Verma:2011yw}
\begin{equation} \label{weak vertex}   
   \begin{split}
\langle\bar{D}^{\ast0}|(c\bar{b})|B^+\rangle=&\epsilon_\nu^\ast\Big\{i\epsilon^{\mu\nu\alpha\beta}P_\alpha q_{\beta}\frac{V(q^2)}{m_{\bar{D}^{\ast0}}+m_{B^+}}-g^{\mu\nu}(m_{\bar{D}^{\ast0}}+m_{B^+})A_1(q^2)+P^\mu P^\nu\frac{A_2(q^2)}{m_{\bar{D}^{\ast0}}+m_{B^+}}\\
    &+q^\mu P^\nu\left[\frac{m_{\bar{D}^{\ast0}}+m_{B^+}}{q^2}A_1(q^2)-\frac{m_{B^+}-m_{\bar{D}^{\ast0}}}{q^2}A_2(q^2)-\frac{2m_{\bar{D}^{\ast0}}}{q^2}A_0(q^2)\right]\Big\},\\  
\langle\bar{D}^{0}|(c\bar{b})|B^+\rangle=&\left[(p_{B^+}+p_{\bar{D}^0})^\mu-\frac{m_{B^+}^2-m_{\bar{D}^0}^2}{q'^2}q'^\mu\right]F_{1D}(q'^2)+\frac{m_{B^+}^2-m_{\bar{D}^0}^2}{q'^2}q'^\mu F_{0D}(q'^2),\\
     \end{split}
\end{equation}
where $P=p_{B^+}+p_{\bar{D}^{\ast0}}$, and $q$ and $q'$ represent the momenta of $D_s^+$ and $D_s^{\ast+}$, respectively.

The form factors of $F_{0D}(t)$, $F_{1D}(t)$, $A_{0,1,2}(t)$, and $V(t)$ with $t\equiv q^{(')2}$ are parameterized as  \cite{Verma:2011yw}
\begin{equation}
    F(t)=\frac{F(0)}{1-a(t/m_B^2)+b(t^2/m_B^4)},
\end{equation}
with $a$ and $b$ being parameters.
 
Combining Eqs. (\ref{vacuum vertex}) and (\ref{weak vertex}), the weak decay amplitudes for $B^+\rightarrow D_s^+\bar{D}^{\ast0}$ and $B^+\rightarrow D_s^{\ast+}\bar{D}^{0}$ can be written as
\begin{equation}
    \begin{split}
        \mathcal{A}(B^+\rightarrow D_s^+\bar{D}^{\ast0})=&\mathcal{M}_\mu(B^+\rightarrow D_s^+\bar{D}^{\ast0})\epsilon^\mu(p_{\bar{D}^{\ast0}}),\\
        \mathcal{A}(B^+\rightarrow D_s^{\ast+}\bar{D}^{0})=&\mathcal{M}_\mu(B^+\rightarrow D_s^{\ast+}\bar{D}^{0})\epsilon^\mu(p_{D_s^{\ast+}}), 
    \end{split}
\end{equation}
with
\begin{equation}
    \begin{split}
        \mathcal{M}_\mu(B^+\rightarrow D_s^+\bar{D}^{\ast0})&=\frac{G_F}{\sqrt{2}}V_{cb}V_{cs}^\ast a_1f_{D_s}\Big\{-q_\mu (m_{B^+}+m_{\bar{D}^{\ast0}})A_1(q^2)+q^\alpha P_\alpha P_\mu \frac{A_2(q^2)}{m_B^+ +m_{\bar{D}^{\ast0}}}\\
       &+P_\mu\left[(m_{B^+}+m_{\bar{D}^{\ast0}})A_1(q^2)-(m_{B^+}-m_{\bar{D}^{\ast0}})A_2(q^2)-2m_{\bar{D}^{\ast0}}A_0(q^2)\right]\Big\},\\    
        \mathcal{M}_\mu(B^+\rightarrow D_s^{\ast+}\bar{D}^{0})=&\frac{2G_F}{\sqrt{2}}V_{cb}V_{cs}^\ast a_1'm_{D_s^{\ast+}}f_{D_s^{\ast+}} p_{\bar{D}^0\mu} F_{1D}(q^{'2}).
    \end{split}
\end{equation}

\subsection{The BS equation for a two pseudoscalar system}
After the weak decay, the final-state strong interactions induce the rescattering of the produced charmed meson pairs into the $D_s\bar{D}_s$ and $D\bar{D}$ bound states. This process involves two ingredients: the strong  decay vertices $\bar{D}^\ast\to\bar{D}_s K$ and $D_s^\ast\to DK$, and the $D_s\bar{D}_s$ ($D\bar{D}$) interaction mediated by light vector-meson exchanges. Based on the heavy-quark sysmmetry and chiral symmetry, the relevant effective Lagrangians are given by~\cite{Cheng:2004ru}
\begin{equation}
\begin{split}
    \mathcal{L}_{D^\ast D \mathcal{P}}&=-ig_{D^\ast D \mathcal{P}}(D_i^\dagger\partial_\mu \mathcal{P}_{ij}D^{\ast\mu}_j-D_i^{\ast\mu\dagger}\partial_\mu\mathcal{P}_{ij}D_j)+\text{H.c.},\\
    \mathcal{L}_{D D \mathcal{V}}&=-ig_{DD\mathcal{V}}D_i^\dagger\overleftrightarrow{\partial}_\mu D_j(\mathcal{V})^\mu_{ij}+\text{H.c.},\\
\end{split}
\end{equation}
where $D^{(\ast)}=(D^{(\ast)0},D^{(\ast)+},D^{(\ast)+}_s)$, $\mathcal{P}$ and $\mathcal{V}$ are $3\times3$ matrices of the pseudoscalar octet and vector meson nonet, respectively, 
\begin{equation}
\label{pseudoscalar}
\mathcal{P}=\left(
\begin{array}{ccc} \frac{\pi^0}{\sqrt{2}}+\frac{\eta}{\sqrt{6}} &\pi^+&K^+\\
                  \pi^-& -\frac{\pi^0}{\sqrt{2}}+\frac{\eta}{\sqrt{6}} &K^0\\
                     K^- &  \bar{K}^0 &    -\sqrt{\frac{2}{3}}\eta  \\
\end{array} \right),
\end{equation}
and
\begin{equation}
\label{vector}
\mathcal{V}=\left(
\begin{array}{ccc} \frac{\omega}{\sqrt{2}}+\frac{\rho^0}{\sqrt{2}} &\rho^+&K^{\ast+}\\
                  \rho^-&\frac{\omega}{\sqrt{2}}-\frac{\rho^0}{\sqrt{2}}&K^{\ast0}\\
                     K^{\ast-} &  \bar{K}^{\ast0} &    \phi  \\
\end{array} \right).
\end{equation}

The flavor wave functions of the isoscalar $D_s\bar{D}_s$ and $D\bar{D}$ systems are taken as
\begin{equation}
    \begin{split}
        |X_{D\bar{D}}\rangle_{I=0}&=\frac{1}{\sqrt{2}}\left(|D^+D^-\rangle+|D^0\bar{D}^0\rangle\right),\\
        |X_{D_s\bar{D}_s}\rangle_{I=0}&=|D_s^+D_s^-\rangle.
    \end{split}
\end{equation}

For these two systems, the BS wave function of a bound state is defined as
\begin{equation}
\begin{split}
\chi_P(x_1,x_2)=&\langle0|T\phi(x_1)\phi^\dag(x_2)|P\rangle\\ 
=&e^{-iPX}\int\frac{d^4p}{(2\pi)^4}e^{-ipx}\chi_P(p),
\end{split}
\end{equation}
where $\phi(x_1)$ and $\phi(x_2)$ are the field operators of the constituent mesons, $P= Mv$ is the total momentum of the bound state, with $v$ its four-velocity, $p$ is the relative momentum, $X = \lambda_1x_1 -\lambda_2x_2$ and $x = x_1 -x_2$ are the center-of-mass and relative coordinates, with $\lambda_{1, 2}=m_{1, 2}/(m_1+m_2)$, where $m_1$ and $m_2$ are the masses of the pseudoscalar meson and the anti-pseudoscalar meson, respectively.  Accordingly, the constituent momenta are $p_1=\lambda_1 P+p$ and $p_2=\lambda_2 P-p$.

By expanding the four-point Green function in terms of  the two-particle irreducible kernel $\bar{K}$, one obtains the homogeneous BS equation,
\begin{equation}\label{BS eq}
    \chi_P(p)=S(p_1)\int\frac{d^4q}{(2\pi)^4}\bar{K}_P(p,p')\chi_P(q)S(p_2),
\end{equation}
together with the normalization condition for the BS wave function,
\begin{equation}
    i\int\frac{d^4pd^4p'}{(2\pi)^8}\bar{\chi}_P(p)\frac{\partial}{\partial P^0}\left[I_P(p,p')+\bar{K}_P(p,p')\right]\chi_P(p')=1,\quad P^0=E_P,
\end{equation}
where $I_P(p,q)=-(2\pi)^4\delta^4(p-q)S^{-1}(p_1)S^{-1}(p_2)$ and $\bar{K}_P(p,q)$ is the sum of all the two-particle irreducible Feynman diagrams. For convenience, we decompose the relative momentum into a longitudinal component along $v$, $p_l\equiv v\cdot p$, and a transverse component, $p_t^\mu\equiv p^\mu-p_l v^\mu$. Then, the constituent propagators can then be written as
\begin{equation}
\begin{split}
    S(p_1)=&\frac{i}{(\lambda_1M+p_l)^2-w_1^2},\\
    S(p_2)=&\frac{i}{(\lambda_2M-p_l)^2-w_2^2},
\end{split}
\end{equation}
with $w_{1,2}=\sqrt{m_{1,2}^2-p_t^2}$.

Based on the effective Lagrangian, the $t$-channel interaction kernel at the tree level in the ladder approximation is given by
\begin{equation}
    \bar{K}_P(p,p')=(2\pi)^4\delta^4(p'_1+p'_2-p_2-p_2)g_{DD\mathcal{V}}^2(p_1+p'_1)^\mu(p_2+p'_2)^\nu\Delta_{\mu\nu}(k),
\end{equation}
where $k=p_1-p'_1$ is the momentum of the exchanged vector meson and $\Delta_{\mu\nu}(k)$ is its propagator.

To account for the internal structure and finite-size effects of the interacting hadrons, a monopole form factor is introduced at each $t$-channel vertex,
\begin{equation}
    F(k^2)=\frac{\Lambda^2-m^2_V}{\Lambda^2-k^2},
\end{equation}
where $m_V$ is the mass of the exchanged meson and $\Lambda$ is the corresponding cutoff. Since the $D\bar{D}$ interaction can proceed through the exchange of mesons with different masses, the cutoff is parametrized as $\Lambda=m_V+\alpha\,\Lambda_{\text{QCD}}$ with $\Lambda_{\text{QCD}}=220$~MeV. The dimensionless parameter $\alpha$, expected to be of order unity, depends on the specific exchanged and external particles. Since it cannot be determined from first principles, it is treated as a phenomenological input.

To simplify the BS equation~\eqref{BS eq}, we further adopt the covariant instantaneous approximation for the kernel by setting $p_l=q_l=0$. In this approximation, the energy exchange between the constituents is neglected while their longitudinal momenta along $v$ remain unchanged. This approximation is justified for the present systems because the binding energies of the $D_s\bar{D}_s$ and $D\bar{D}$ systems are much smaller than the masses of the charmed-strange and charmed mesons. Under this approximation, the kernel reduces to $\bar{K}_P({p}_t,{q}_t)$.

\subsection{Production amplitudes for $B^+\to XK^+$}
Combining the weak-decay amplitudes with the BS wave functions of the bound states, we obtain the production amplitudes for $B^+\to X_{s\bar{s}}K^+$ and $B^+\to X_{q\bar{q}}K^+$ as
\begin{equation}
    \begin{split}
        \mathcal{A}_{B^+\rightarrow X_{ss}K^+}=&\int\frac{d^4p}{(2\pi)^4}g_{\bar{D}^{\ast}D_sK}\mathcal{M}_\mu(B^+\rightarrow D_s^+\bar{D}^{\ast0})p_{K^+\nu}\frac{-i\left(g^{\mu\nu}-p_{\bar{D}^{\ast0}}^\mu p_{\bar{D}^{\ast0}}^\nu/m_{\bar{D}^{\ast0}}^2\right)}{p_{\bar{D}^{\ast0}}^2-m_{\bar{D}^{\ast0}}^2+im_{\bar{D}^{\ast0}}\Gamma_{\bar{D}^{\ast0}}}\chi_P(p),\\
        \mathcal{A}_{B^+\rightarrow X_{qq}K^+}=&\int\frac{d^4p}{(2\pi)^4}g_{\bar{D}_s^{\ast}DK}\mathcal{M}_\mu(B^+\rightarrow D_s^{\ast+}\bar{D}^{0})p_{K^+\nu}\frac{-i\left(g^{\mu\nu}-p_{D_s^{\ast+}}^\mu p_{D_s^{\ast+}}^\nu/m_{\bar{D}^{\ast0}}^2\right)}{p_{D_s^{\ast+}}^2-m_{D_s^{\ast+}}^2+im_{D_s^{\ast+}}\Gamma_{D_s^{\ast+}}}\chi_P(p).\\
    \end{split}
\end{equation}

With these amplitudes, the differential decay width of the $B^+$ meson is given by
\begin{equation}
    d\Gamma=\frac{1}{32\pi^2}|\mathcal{M}|^2\frac{|\vec{p}_{K^+}|}{m_B^2}d\Omega,
\end{equation}
where $\vec{p}_{K^+}$ is the momentum of the $K^+$ meson in the rest frame of the $B^+$ meson and $d\Omega=\sin\theta d\theta d\phi$ is the solid angle of $K^+$.

\section{Numerical results}
\label{NR}
In this work, the BS wave functions are employed to solve out the internal dynamics of the $D_s\bar{D}_s$ and $D\bar{D}$ bound states and to investigate their effects when applying them to the production rates of these bound states in $B$ decays. The formation of the $D_s\bar{D}_s$ and $D\bar{D}$ bound states depends sensitively on the couplings $g_{HHV}$, with $H=D,D_s$ and $V=\rho,\omega,\phi$. For the $D_s\bar{D}_s$ system, the relevant coupling is $g_{D_sD_s\phi}$, whereas for the $D\bar{D}$ system the relevant couplings are $g_{DD\rho}$ and $g_{DD\omega}$. These couplings can be estimated from the parameter $\beta$ associated with heavy-quark and chiral symmetries \cite{Casalbuoni:1996pg,Cheng:2004ru}, or extracted from nonperturbative approaches such as lattice QCD \cite{Can:2012tx} and QCD sum rules \cite{Ball:2004rg}. In the present analysis, we adopt the results of Ref. \cite{Jiang:2024equ}, in which the couplings were evaluated in a unified QCD light-cone sum-rule framework including improved corrections beyond the leading order in $\alpha_s$, The resulting values are 
\[
g_{D_sD_s\phi}=3.650^{+0.93}_{-0.59},\qquad
g_{DD\rho}=4.30^{+0.82}_{-0.72},\qquad
g_{DD\omega}=2.80^{+0.54}_{-0.48}.
\]
For numerical convenience, we define three input sets, denoted as Set A, Set B, and Set C, corresponding to the lower, central, and upper values of the relevant couplings, respectively.

After fixing the couplings, the model contains only one free parameter, the cutoff $\Lambda$, which is parameterized as $\Lambda=m_V+\alpha \Lambda_{\rm{QCD}}$, where $\alpha$ is a dimensionless parameter of order unity, constrained by the typical hadronic size scale. To search for possible bound-state solutions, we vary $\alpha$ in the range [0.5, 9]. The binding energy is defined as $E_b = m_1 + m_2 - M$, and the $D_s\bar{D}_s$ and $D\bar{D}$ systems are treated as shallow bound states with $E_b\in[1, 30]$ MeV. For a given binding energy, the corresponding value of $\alpha$ is then obtained by solving the BS equation.

For the weak decay inputs, we take $G_F=1.166\times10^{-5}$ $\text{GeV}^{-2}$, $V_{cb}=0.041$, $V_{cs}=0.987$, $f_{D_s}=250$ $\text {MeV}$, and $f_{D_s^\ast}=272$ MeV, following Ref. \cite{ParticleDataGroup:2024cfk}. The transition form factors are adopted from the covariant light-front quark model: $(F_{1D}(0), a, b)^{B^+\rightarrow \bar{D}^0}=(0.67, 1.22, 0.36)$, $(A_0(0), a, b)^{B^+\rightarrow \bar{D}^{\ast0}}=(0.68, 1.21, 0.36)$, $(A_1(0), a, b)^{B^+\rightarrow \bar{D}^{\ast0}}=(0.65,0.60, 0.00)$, and $(A_2(0), a, b)^{B^+\rightarrow \bar{D}^{\ast0}}=(0.61,1.12, 0.31)$. The effective Wilson coefficients are fixed as $a_1=0.93$ and $a'_1=0.81$ \cite{Xie:2022lyw}, which are extracted from the measured branching fractions of $B^+\rightarrow D_s^+\bar{D}^{\ast0}$ and $B^+\rightarrow D_s^{\ast+}\bar{D}^{0}$. The couplings $g_{D_s^+\bar{D}^{\ast0}K^+}$ and $g_{{D}_s^{\ast+}D^{0}K^+}$ are determined using SU(3)-flavor symmetry, $g_{D_s^+\bar{D}^{\ast0}K^+}=g_{{D}_s^{\ast+}D^{0}K^+}=\sqrt{2}g_{D^{\ast0}D^0\pi}$, where $g_{D^{\ast0}D^0\pi}$ is extracted from the measured decay width $D^{\ast0}\rightarrow D^0\pi$ \cite{ParticleDataGroup:2024cfk}. Since the main purpose of this work is to examine how the production rates depend on the BS wave functions of the $D_s\bar{D}_s$ and $D\bar{D}$ bound states, the uncertainties of these input parameters are not further considered.

\begin{figure}[h]
\centering
    \includegraphics[width=0.6\textwidth]{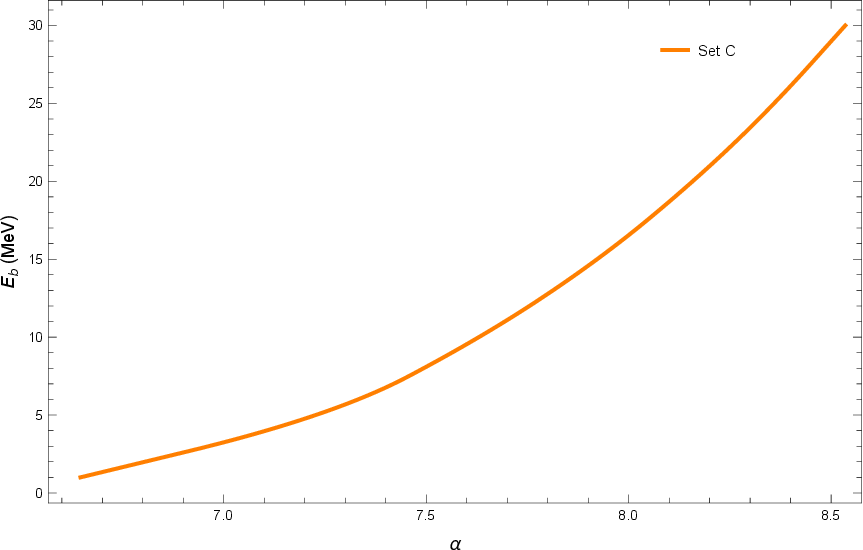}
    \caption{Values of $\alpha$ and $E_b$ for the possible bound states for the $D_s\bar{D}_s$ system.}
  \label{fig3}
\end{figure}

\begin{figure}[h]
\centering
    \includegraphics[width=0.6\textwidth]{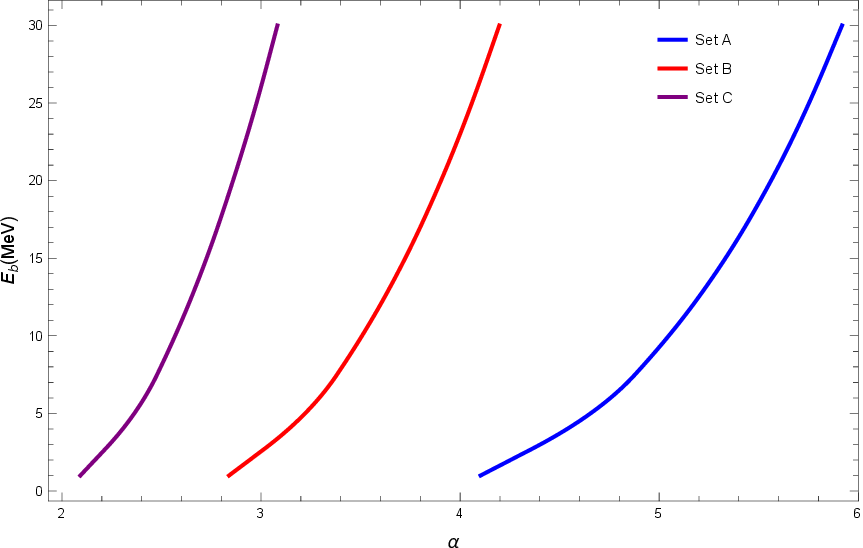}
    \caption{Values of $\alpha$ and $E_b$ for the possible bound states for the $D\bar{D}$ system.}
  \label{fig4}
\end{figure}

The numerical results for the possible $D_s\bar{D}_s$ and $D\bar{D}$ bound states are shown in Figs. \ref{fig3} and \ref{fig4}, respectively. For the $D_s\bar{D}_s$ system, Sets A and B do not support bound state solutions, since the required values of $\alpha$ are too large to be phenomenologically acceptable. Only when $g_{D_sD_s\phi}$ takes its upper value (Set C) does a bound state solution emerge, with $\alpha\in[6.64, 8.35]$. Values of $\alpha$ much larger than unity are generally considered less natural in this phenomenological parametrization. Therefore, for the $D_s\bar{D}_s$ system, the large values of $\alpha$ required to produce even a shallow bound state indicate that binding is not easily achieved  through the $\phi$ exchange interaction alone. In principle, additional exchanges, such as those mediated by $\phi(1680)$ and $f_0(980)$ (the contribution from $\sigma$ exchange is very small \cite{Li:2022shq,Ding:2008gr}), may also contribute to the $D_s\bar{D}_s$ interaction. However, since their couplings to $D_s$ mesons are presently poorly known, these contributions are not included in the present calculation. 

For the $D\bar{D}$ system, bound state solutions are found for all the three coupling sets. The corresponding ranges of $\alpha$ are [4.10,5.92], [2.84,4.20], and [2.09,3.08] for Sets A, B, and C, respectively. Compared with the $D_s\bar{D}_s$ system, the significantly smaller values of $\alpha$ required for $D\bar{D}$ bound states indicate a stronger effective attraction in this system. This suggests that the $D\bar{D}$ interaction is more favorable for forming a shallow bound state within the present framework. In addition, the noticeable shift of the allowed $\alpha$ interval among the three coupling sets shows that the $D\bar{D}$ solutions are quantitatively sensitive to the input couplings.

\begin{figure}[ht]
\centering
    \includegraphics[width=0.6\textwidth]{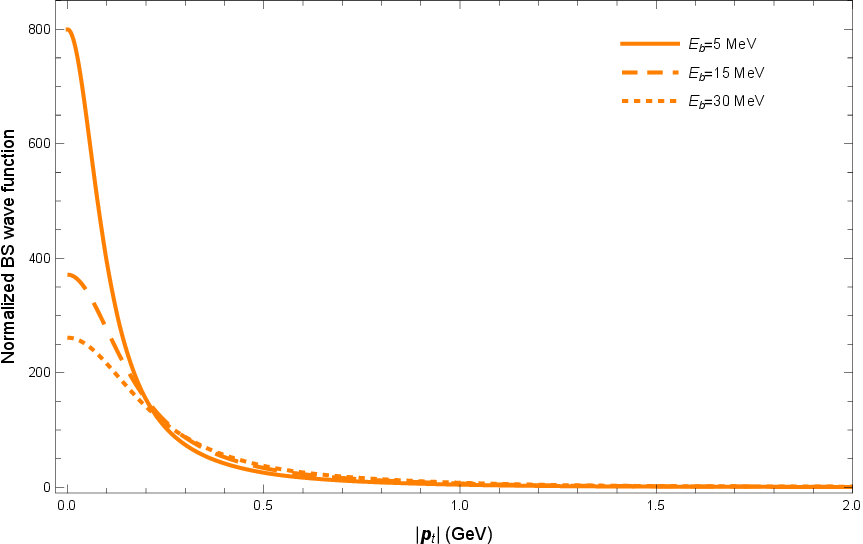}
    \caption{Normalized BS wave function for the $D_s\bar{D}_s$ bound state.}
  \label{fig5}
\end{figure}

\begin{figure}[htb]
\centering
\includegraphics[width=0.6\textwidth]{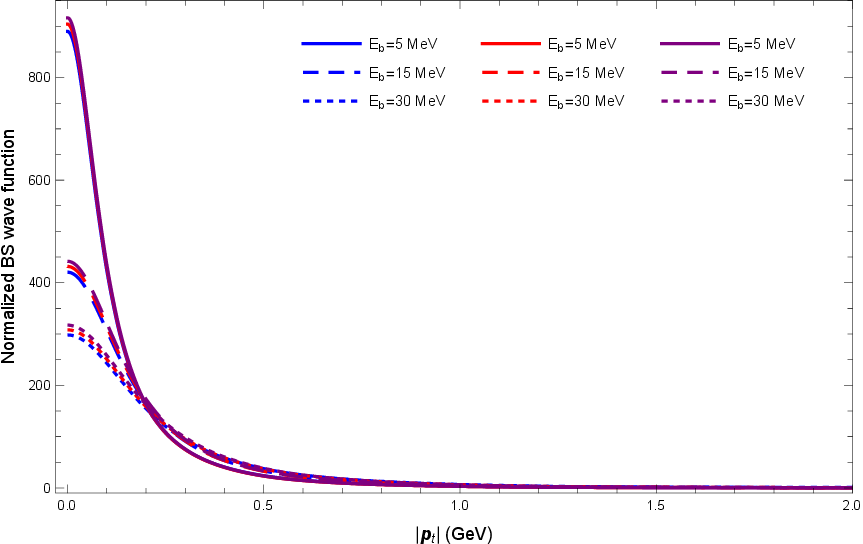}
\caption{Normalized BS wave functions for the $D\bar{D}$ bound state. The blue, red, and purple curves correspond to Sets A, B, and C, respectively.}
\label{fig6}
\end{figure}

The normalized scalar BS wave functions for the $D_s\bar{D}_s$ and $D\bar{D}$ bound states are shown in Figs. \ref{fig5} and \ref{fig6}, respectively, for three representative binding energies, $E_b$ = 5, 15, and 30 MeV. As seen from both figures, as the binding energy increases, the peak value of the BS wave functions at low-momentum region decreases significantly.  Fig. \ref{fig6} further shows that, for a fixed binding energy, the normalized BS wave functions of the $D\bar{D}$ bound state are insensitive to the choice of the coupling set, since the results for Sets A, B, and C are very close to each other.

\begin{figure}[ht]
\centering
    \includegraphics[width=0.6\textwidth]{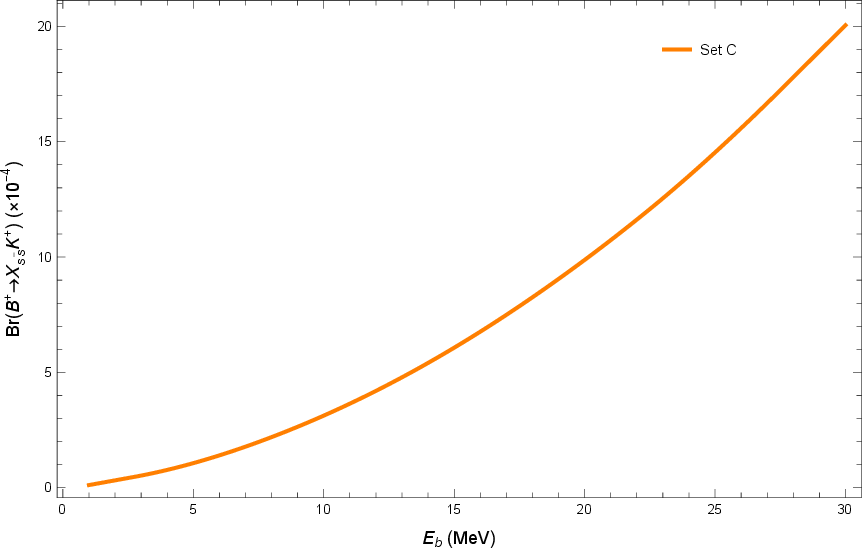}
    \caption{Branching ratios of $B^+\rightarrow X_{s\bar{s}}K^+$ with $X_{s\bar{s}}$ as the possible $D_s\bar{D}_s$ bound state.}
  \label{fig7}
\end{figure}

\begin{figure}[ht]
\centering
    \includegraphics[width=0.7\textwidth]{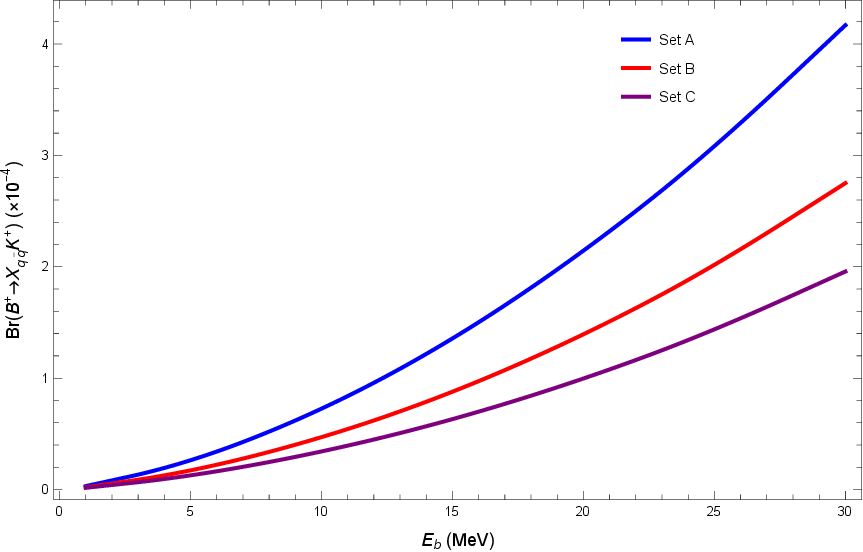}
    \caption{Branching ratios of $B^+\rightarrow X_{q\bar{q}}K^+$ with $X_{q\bar{q}}$ the possible $D\bar{D}$ bound state. The blue, red, and purple curves correspond to Sets A, B, and C, respectively.}
  \label{fig8}
\end{figure}

Since the BS wave function $\chi_P(p)$ is Lorentz covariant, it can be directly incorporated into the decay amplitudes of the processes $B^+\rightarrow X_{s\bar{s}}K^+$ and $B^+\rightarrow X_{q\bar{q}}K^+$. Because $\chi_P(p)$ includes the internal dynamics of the bound states, the corresponding branching ratios are sensitive to the BS wave functions of the $D_s\bar{D}_s$ and $D\bar{D}$ systems, which are identified here with $X_{s\bar{s}}$ and $X_{q\bar{q}}$, respectively. The resulting branching ratios as functions of the binding energy are displayed in Figs. \ref{fig7} and \ref{fig8}. For the $D_s\bar{D}_s$ bound state, corresponding to the process $B^+\rightarrow X_{s\bar{s}}K^+$, the branching ratio shown in Fig. \ref{fig7} ranges from $1.09\times10^{-5}$ to $20.06\times10^{-4}$. If $X(3915)$ is interpreted as a $D_s\bar{D}_s$ bound state and its mass is taken as $M_{X(3915)}=3922.1$ MeV from the PDG \cite{ParticleDataGroup:2024cfk}, the predicted branching fraction is $\mathcal{B}(B^+\rightarrow X(3915)K^+)=5.79\times10^{-4}$. Although this value is somewhat above the PDG upper limit $2.8\times10^{-4}$, it remains comparable to other theoretical estimates based on the same $D_s\bar{D}_s$ molecular interpretation, such as $(9\pm2.5)\times10^{-4}$ in Ref. \cite{Xie:2022lyw} and $6\times10^{-4}$ in Ref.~\cite{Li:2015iga}. 

For the $D\bar{D}$ bound state, the branching fraction for $B^+\rightarrow X_{q\bar{q}}K^+$ is found to lie in the ranges $3.05\times10^{-6}$ to $4.14\times10^{-4}$, $2.05\times10^{-6}$ to $2.75\times10^{-4}$, and $1.56\times10^{-6}$ to $1.96\times10^{-4}$, for Sets A, B, and C, respectively. We find that the branching fraction decreases as the couplings increase, while its dependence on the coupling set becomes weaker for a smaller binding energy. This behavior reflects the interplay between the interaction kernel of the bound state and the normalization of the BS wave function. If the $D\bar{D}$ bound state decays predominantly into $\eta_c\eta$, the process $B^+\rightarrow\eta_c\eta K^+$ may provide an experimentally accessible probe of this molecular configuration. The current experimental upper limit, $\mathcal{B}(B^+\rightarrow\eta_c\eta K^+)< 2.2\times10^{-4}$ can therefore be used as a useful reference.  In the effective Lagrangian approach of Ref. \cite{Xie:2022lyw}, the branching fraction for $B^+\rightarrow X_{q\bar{q}}K^+$ was estimated to be in the range $(1.3\pm0.4)\times10^{-4}$ to $(5.2\pm1.5)\times10^{-4}$, which is somewhat larger than our present results. For the $D\bar{D}$ bound state with a binding energy $E_b=4$~MeV, as suggested by lattice QCD~\cite{Prelovsek:2020eiw}, we obtain  $\mathcal{B}(B^+\rightarrow X_{q\bar{q}}K^+)$ = $0.19\times10^{-4}$, $0.12\times10^{-4}$, and $0.10\times10^{-4}$ for Sets A, B, and C, respectively.

\section{Conclusions}

In this work, we have investigated the production of possible $D_s\bar{D}_s$ and $D\bar{D}$ bound states in $B$ decays within the BS framework. The BS equations for the two heavy meson systems were solved in the one-boson-exchange model, and the resulting BS wave functions were incorporated into the weak decay amplitudes to evaluate the branching fractions for $B^+\to X_{q\bar q}K^+$ and $B^+\to X_{s\bar s}K^+$.

Our numerical analysis shows that the two systems exhibit markedly different binding behaviors. For the $D_s\bar{D}_s$ system, a bound state solution can be obtained only for the largest coupling set, and relatively large values of the cutoff parameter $\alpha$ are required. This indicates that, within the present framework, the attraction generated by $\phi$ exchange alone is not strong enough to naturally produce a shallow $D_s\bar{D}_s$ bound state. In contrast, for the $D\bar{D}$ system, bound-state solutions exist for all the three coupling sets with considerably smaller values of $\alpha$, suggesting that this system is more favorable for binding in the present model.

The corresponding normalized BS wave functions were obtained for representative binding energies and were then used to calculate the production rates in $B$ decays. For the $D_s\bar{D}_s$ bound state, the predicted branching fraction of $B^+\to X_{s\bar s}K^+$ ranges from $1.09\times10^{-5}$ to $20.06\times10^{-4}$. If $X(3915)$ is interpreted as a $D_s\bar{D}_s$ bound state, the corresponding branching fraction is found to be $5.79\times10^{-4}$. For the $D\bar{D}$ bound state, the branching fraction of $B^+\to X_{q\bar q}K^+$ is predicted to lie in the range from $1.56\times10^{-6}$ to $4.14\times10^{-4}$, depending on the binding energy and the choice of the coupling set.

These results indicate that $B$ meson decays may provide a useful production environment for searching for heavy meson molecular states. In particular, within the present BS framework, the $D\bar{D}$ system appears to be a more promising bound state candidate than the $D_s\bar{D}_s$ system. Further improvements in constraining the relevant heavy meson couplings and in incorporating additional interaction mechanisms would be valuable for enhancing the quantitative reliability of the present predictions.
	
\section{Acknowledgments}
This work was supported by the National Natural Science Foundation of China (No. 12405115, No. 12105149, No. 12475096, and No. 12275024), the High-Level Scientific Research Fund of Ningbo University under Grants No. GJPY2026032.

\clearpage
\bibliography{ref.bib}
\end{document}